\documentclass[11pt]{article}

\def\myatitle{C. F. von Weizs\"acker's Reconstruction of Physics:}
\def\mybtitle{Yesterday, Today, Tomorrow}

\def\be{\begin{equation}}
\def\ee{\end{equation}}
\def\bea{\begin{eqnarray}}
\def\eea{\end{eqnarray}}
\def\ba{\begin{array}}
\def\ea{\end{array}}
\usepackage{amssymb,bbm}  
\def\setR{\mathbbm R}     
\def\setC{\mathbbm C}     
\def\set1{\mathbbm 1}
\def\setS{\mathbb S}      

\def\quabla{\raisebox{-.2ex}{\Large$\square$}}

\begin{document}
\sloppy

\thispagestyle{empty}

\vspace*{-20mm}

\noindent \textit{In:
L. Castell and O. Ischebeck (eds.), 
``Time, Quantum and Information (Essays in Honor of C. F. von Weizs\"{a}cker),''
Springer, Berlin, 2003.}

\vspace{10mm}

\begin{center}

{\huge \myatitle}\\[3mm]
{\huge \mybtitle}

\vspace{10mm}

{\Large Holger Lyre}

\vspace{3mm}

{\small
Philosophy Department,\\ 
University of Bonn\\
}

\vspace{15mm}

\end{center}


 \hfill \parbox{12cm}{\em\footnotesize
 Heisenberg ... hatte seine nichtlineare Spinorfeldtheorie
 auf die Voraussetzung der allgemeinen ... Quantentheorie und die
 Forderung der Poincar\'e-Gruppe  und der Isospingruppe gegr\"undet.
 Als ich ihn fragte, warum genau diese Gruppen, sagte er etwa:
 `Das kann ich nicht mehr begr\"unden.
 Mit \textit{einer} Forderung mu\ss{} man eben anfangen.'
 Ich fragte: `Warum \"uberhaupt Gruppen?' Er, etwa:
 `Wenn man mit einer Forderung anf\"angt, dann ist
 Symmetrie der beste Anfang. Symmetrie ist sch\"on,
 das wu\ss{}te schon Platon.
 Darin dr\"uckt sich die zentrale Ordnung aus.' ...

 Ich aber blieb unbefriedigt. Mir schien, man solle eine rationale
 Behauptung wie die Geltung einer Symmetrie wom\"oglich noch rational
 begr\"unden. Meine Vermutung war: Symmetrie bedeutet die Trennbarkeit
 des jeweils untersuchten Gegenstandes vom Rest der Welt. ...
 Die Frage ist, ob man diesem Gedanken noch eine strengere Fassung
 geben kann.}

 \hfill C. F. von Weizs\"acker (1992, p.~910)\\ 


\section{The Origin: The Philosophy of Physics}

Carl Friedrich von Weizs\"{a}cker's thinking has always crossed
the borders between physics and philosophy. Being a physicist
by training he still feels at home in the physics community,
as a philosopher by passion, however, his mind cannot
stop thinking at the limits of physics. His physical ideas
are based on the general conceptual and methodological preconditions
of physical theories. The above quotation does indicate this:
Heisenberg's hint on symmetry as a natural starting point for
a physicist---even though already philosophically motivated by
Plato---left Weizs\"acker ``dissatisfied.'' He even wanted to
explore the reasons for the symmetries themselves.

Such a line of reasoning about the foundations of physics has
brought Weizs\"acker into an abstract program of a possible
reconstruction of physics in terms of yes-no-alternatives, which
he called ``ur-theory.'' I shall start this paper with a review of
the basic ideas of ur-theory: the definition of an ur and the
connection between ur-spinors and spacetime (Sect.\ \ref{yesterday}:
``Yesterday''). I then go over to some of ur-theory's present
borders: the construction of quantized spacetime tetrads and the
difficulties to incorporate gravity and gauge theories (Sect.\
\ref{today}: ``Today''). It goes without saying that my brief review 
will be far from being complete---see in particular the developments 
presented in Castell et~al. (1975-1986). In the last section (Sect.\
\ref{tomorrow}: ``Tomorrow'') I shall discuss the possible
prospects of ur-theory---partly with a view to modern quantum
gravity approaches, but mainly in connection with its
philosophical implications. Here, one of the crucial questions is,
whether form, or, modern, information is an entity \textit{per se}
and what particular consequences this may have.


\section{Yesterday: Urs, Spinors and Spacetime}
\label{yesterday}

Weizs\"acker's first book on the philosophy of physics,
\textit{Zum Weltbild der Physik}, appeared in 1943 in its first
edition---when his professional main concern still was `real
physics.' But already at this time his `real interest' had shifted
to the foundations of quantum mechanics. About ten years later, at
a spa in Bad Wildungen in 1954, Weizs\"acker had the crucial idea:
the quantum theory of a binary alternative \textit{is} the theory
of objects in a three-dimensional space. Mathematically, he had
just stumbled about the fact that $SU(2)$, the quantum theoretical
symmetry group of a binary alternative, is locally isomorphic to
the three-dimensional rotation group $SO(3)$ in Euclidean space.
Philosophically, however, this was more ``satisfying'' than just
to start with a certain symmetry, since now the symmetry itself
was distinguished by the fact that it is the symmetry of the
simplest logical object---a yes-no-alternative. Couldn't this be a
reason for the three-dimensionality of position space?

Weizs\"acker's approach rests on two main `ingredients':
firstly, the idea of reducing physics to binary alternatives and,
secondly, the connection between spin structures and the structures
of time and space. This second motive has caused
David Finkelstein (1994) 
to subsume Weizs\"acker's approach under the heading of ``spinorism''
and this will be my main concern in this and the following section.
For a moment, however, we shall consider the first mentioned,
rather philosophical ingredient. Here the idea is
that---in its core---physics reduces to predicting measurement
outcomes. Measurement outcomes may be restated
in terms of empirically decidable n-fold alternatives, and,
trivially, any n-fold alternative can be embedded into a
(Cartesian) product of binary alternatives. Binary alternatives,
in turn, can be considered as bits of information. Thus physics
reduces to information or, more precisely, \textit{potential} information.

Weizs\"acker has called the empirically decidable binary alternatives
in his reconstruction \textit{ur-alternatives}
(from the German prefix `Ur-': `original,' `elementary,' `pre-').
A bit whimsically, he later called them \textit{urs}.
Urs may be considered the fundamental objects in physics.
As a matter of principle, any physical object can be built out of them.
The reader may note that at this point an interesting conceptual shift
has occurred. We started from the notion of information, a subjective notion
in the first place, and applied it to the objective notion of a physical object.
This is another way of saying that objects are reduced or even `made out of'
information. It seems that via this shift information has been gained
the status of a substance. In the last section I shall indeed come back to
this challenging formula, as our present working hypothesis, however,
we will take an information-theoretic reductionist view of physics
in the sense that physical objects are entirely characterized by the
information which can be gained from them.

In quantum theory in particular, this view has a lot of plausibility.
Quantum objects are represented in terms of their Hilbert state spaces,
their quantum states correspond to empirically decidable alternatives.
Any quantum object may further be de-composed or embedded into the
tensor product of two-dimensional objects, nowadays called quantum bits
or qubits. Urs, therefore, are in fact nothing but qubits
(cf.\ Sect.\ \ref{tomorrow}).

Weizs\"acker's vision of the structure of physics as a quantum theory
of ur-alternatives has its roots in the above mentioned fact
that the essential symmetry group of urs, which is $SU(2)$,
is the double covering of the rotational symmetry group of three-dimensional
position space. The guiding line is that, if the idea of urs as
fundamental entities is true, the symmetry of such urs should
play an essential role in the reconstruction of physics and in the
phenomenology of our empirical world. And, of course, position space
is probably the most essential feature of the empirical world!

But this only gives rise to spinorism in general, the mathematical motive
of deriving the spacetime structure from a primarily given spin structure.%
\footnote{To the best of my knowledge, Weizs\"acker presented the first hint
on spinorism in a short note from 1951 on the occasion of Werner Heisenberg's
50th birthday, where he drew attention to the fact that both position space
and Hilbert space are provided with a quadratic metric and that in order
to give a deeper reason for this one presumably needs a property common to both
spaces (which indeed could be the spin structure). He closes that perhaps
the question of the structure of the state space is fundamental, whereas
the structure of position space is a dependent and derived one---a rather
bold and visionary remark in 1951 (and, of course, still today)!
If we now take the publication date of this note (Weizs\"acker 1952), 
then this Festschrift can simultaneously be considered as celebrating 50 years
of Weizs\"acker's idea of spinorism. However, the rather mathematical hint from 1951/52,
the second ingredient of ur-theory, was only filled with conceptual life
two years later at the above mentioned spa. This little episode also sheds light
on the crucial interplay between philosophy and physics in Weizs\"acker's thinking.}
A more precise ur-theoretic ansatz was chosen in Weizs\"acker's
1958 papers \textit{Die Quantentheorie der einfachen Alternative
(Komplementarit\"at und Logik II)}---internally called ``KL II''---and
the ``Dreim\"annerarbeit'' KL III together with Erhard Scheibe
and Georg S\"u\ss{}mann. 
KL II starts from the well-known and deep connection between
tensors and spinors: there always exists a mapping between
tensors of order $n$ and spinors of order $2n$.
A lightlike four-vector, in particular, can be written in terms
of the Pauli matrices $\sigma_\mu$ as
\be
\label{kmu}
k_\mu = \sigma^{\mu}_{\dot A B} u^{\dot A} u^B ,
\ee
where $u^A$ is a spinor and dotted indices denote complex conjugate components.
This relation highlights the link between the homogeneous Lorentz group $SO(1,3)$
and  $SL(2,\setC)$, the unimodular group in spinor space
(the spinors do not obey a unitary norm here).
But instead of elaborating the thus defined spin structure of Minkowski space,
the focus of KL III lies on Weizs\"acker's idea of ``multiple quantization.''
Generally speaking, the procedure of quantization can be split into two steps:
(i) the transition from a (classically) discrete number of degrees of freedom
    to infinitely many degrees, and
(ii) the transition to operator-valuedness and, hence, appropriate
    commutation relations.
Starting from a simple, classical yes-no-alternative $a_A$, step (i) means
the construction of a wavefunction $\phi(a_A)$, i.e.\ a spinor $u_A \equiv \phi(a_A)$.
Relation (\ref{kmu}) allows to transform $u_A \Leftrightarrow k_\mu$,
then (ii) implies the transition $k_\mu \to \hat k_\mu$.
This is the `first quantization' of a binary alternative.
On the level of `second quantization' wavefunctions $\varphi(k_\mu)$ occur,
where operators $\hat k_\mu$ act upon. Interpreting $k_\mu$ as energy-momentum
four-vector, we get quantum mechanical wavefunctions $\psi(x_\mu)$ in Minkowski space
after Fourier transformation. The next, third level of quantizing
$\psi(x_\mu)$ corresponds to usual quantum field theory.

Weizs\"acker, Scheibe and S\"u\ss{}mann discovered in their paper the Weyl 
and---after doubling the state space of urs and working with bi-spinors---the
Dirac and Klein-Gordon equation and also the homogeneous Maxwell equations
as algebraic identities!%
\footnote{Details of the derivation must be left out due to lack of space,
the reader may consult KL III or Lyre (1998a, chap. 2.4).}
%
Since for a nullvector one obtains $k_\mu k^\mu = 0$,
the massless Klein-Gordon equation follows for instance from
\be
\label{KG}
\hat k_\mu \hat k^\mu \ \varphi(k_\mu) = 0
\quad \stackrel{FT}{\Longleftrightarrow} \quad \quabla\, \psi(x_\mu) = 0.
\ee
Let $D(\frac{1}{2},0)$ be the fundamental representation of $SL(2,\setC)$ 
and $D(0,\frac{1}{2})$ its complex-conjugate, then the transition to bi-spinors 
effectively means to work with $D(\frac{1}{2},0) \otimes D(0,\frac{1}{2})$. 
I shall show in the next section how this combines 
with a proper usage of the full spin structure of urs.

We have so far presented ur-theoretic arguments to introduce
Minkowski spacetime as a local spacetime model, we shall now
go over to global spacetime considerations.
We start again with a quantum theory of binary alternatives,
but this time with usual quantum bits obeying a unitary norm.
The symmetry group of urs then contains $SU(2)$, $U(1)$ and
the complex conjugation. As a Lie group manifold, this yields
\be
SU(2) \times U(1) \ = \ \setS^3 \times \setS^1 .
\ee
Weizs\"acker (1985) has made the far-reaching assumption---for
which the above portrayed spinorism is just the motive---that $\setS^3$
itself should be considered a model of global cosmic space. 
It follows that urs are wavefunctions on $\setS^3$.
As Thomas G\"ornitz (1988) 
has shown, this leads to a remarkably fresh look at the large numbers
in physics on the basis of the multiplicity of the regular representations
of $SU(2)$, in accordance with the Planck scale and Weizs\"acker's
earlier calculations. This will be explained in a moment.
Moreover, for the appropriate treatment of global cosmic time
and time in general, the distinction between past and future
plays a major role in Weizs\"acker's philosophy of physics.
He considers it a very precondition of empirical science, 
on which the concept of separable alternatives has to be built.
In my opinion it is therefore highly implausible to take $U(1)$ 
as a model of cosmic time, which would be cyclic then.
Obviously, time demands a special treatment
(see, however, Castell 1975 for an opposite view). 

We now seek to calculate the number of urs, starting from Weizs\"acker's
early estimations in the 60's. As a wavefunction on global space,
one ur can be thought of as the alternative of being
in the `one half' or the `other half' of the universe. Suppose now
we want to localize a nucleon in the universe. The Compton wavelength
can be understood as a natural measure of localizing a certain particle.
For a nucleon, the ratio between the cosmic radius $R$ and the Compton wavelength
$\lambda$ is about $10^{40}$. We therefore have to decide $10^{40}$ alternatives
in each spatial direction to localize that nucleon. Hence, in ur-theory the assumption
is made that a nucleon \textit{is} $10^{40}$ urs (up to two or three orders of magnitude).
Accordingly, an electron is about $10^{37}$ urs.

The Compton wavelength $\lambda$ of a nucleon is actually
a distinguished measure of length.
If we take the whole energy content of the universe to subdivide
space into equal intervals, then $\lambda$ drops out.
In this sense it is indeed an elementary length.
We may of course measure up smaller regions of space, but then,
as a matter of principle, do we loose the possibility to perform
measurements of other regions at the same time with the same accuracy.
In this sense the number of elementary spatial cells,
which is $N=\left( \frac{R}{\lambda} \right)^3=10^{120}$,
is the total number of elementary ``events'' in the world and,
thus, the total number of urs (the one bit decision whether,
for instance, a particular cell is occupied by a nucleon).
Accordingly, the dimension of the Hilbert space of urs is $2^{{10}^{120}}$.
As a \textit{result}, not as an input, Weizs\"acker got the number of nucleons
as $\frac{10^{120}}{10^{40}}=10^{80}$---in accordance with empirical results!

As already mentioned, G\"ornitz has refined these estimations
and put them on a more solid, group-theoretical basis.
He considered the regular representation of $SU(2)$, i.e.\ the representation
in terms of square-integrable functions of the Hilbert space ${\cal L}^2(\setS^3)$
on $SU(2)$ itself as a homogeneous space $\setS^3$.
The multiplicity of the irreducible representations of the reduced tensor product
of spin-${1 \over 2}$-representations of $SU(2)$ shows a characteristic
cut-off at functions with a wavelength of the order $l_o = \frac{R}{\sqrt{N}}$,
where $R$ is the radius of $\setS^3$ and $N$ the number of ur-functions
(cf.\ G\"ornitz 1988 for details).
Obviously, $l_o$ is now the lower bound of spatial distances which can be measured
in a cosmos with $N$ urs at maximum. $l_o$ is indeed a fundamental elementary length
in the sense that it represents the smallest spatial resolution physically possible.
Remarkably, from the above $N=10^{120}$ we get $l_o=10^{-60} R$,
i.e. the Planck length!
Conversely, if we already know about the Planck scale from other
considerations (as we usually do), we get $N=10^{120}$ as a result---in
accordance with Weizs\"acker's original, more hand-waving estimations.


\section{Today: Tetrads, Gravity and Gauges}
\label{today}

In the preceding section we have discussed the possibility of introducing
spacetime from an ur-hypothesis and also deriving the form of the
basic equations for matter fields (mainly the Dirac equation)
and the simplest free interaction field (the homogeneous Maxwell equations).
From a broader perspective two questions arise:
Shouldn't the ur-theoretic ansatz, a way of deriving spacetime itself,
directly lead to gravity (or, perhaps, even quantum gravity)?
Shouldn't we be able to derive interaction gauge theories in general?
Ambitious questions, indeed, but it is perhaps not astonishing
that in abstract accounts like ur-theory questions like these arise
at the very beginning. And despite of their ambitious character and
also far from really answering them I try to discuss
in the following some possible ways to tackle these questions.

As a working hypothesis, let us, again, take $SU(2)=\setS^3$ as a model
of global position space. Urs can be understood as non-local functions 
on $SU(2)$ and do naturally represent a spinor dyad 
(with spinors $u^A$, $v^A$ satisfying $u_A v^A = - v_A u^A = 1$).
It is now well-known that a spinor dyad is equivalent to a null-vector
vierbein or tetrad, where the four null-vectors have the form (\ref{kmu}),
but consist of mixed combinations of $u_A$ and $v_A$
(the details of this and the following paragraph can be found in 
Lyre 1998b). 
By considering suitable linear combinations of the null-vectors,
such a null-tetrad can generally be written in real-valued form
$\theta_\mu^\alpha=(t_\mu, x_\mu, y_\mu, z_\mu)$,
where the space-like vectors $x^\mu, y^\mu, z^\mu$ represent 
a spatial dreibein tangent to $\setS^3$ 
together with an orthogonal time-like vector $t^\mu$.
The interesting point is that, since the tetrad is written in terms
of ur-spinor components, a (first) quantization of urs also induces
a quantization of the tetrad. Such a quantized ur-tetrad, however,
means nothing but quantized coordinates in our spacetime model.

As a remarkable of the quantized ur-tetrad it turns out that 
the time operator $\hat t_\mu$ in is just the ur-number operator
$\hat n = \frac{1}{2} \sum_r \{ \hat a_r^+, \hat a_r \}$
in the Fock space of Bose urs.
This is consistent with Weizs\"acker's assumption that the growth
of the total number of urs is a measure of temporal cosmic evolution.
Now, since the number operator has indeed a lower bound at zero,
the global spacetime model $\setS^3 \times \setR^+$ with a time parameter
manifold $\setR^+$ seems justified (thus avoiding globally closed
time-like curves as mentioned in Sect.\ \ref{yesterday}).

Can all this give us a hint how to describe (quantum) gravity in ur-theory?
We may think of $\theta_\mu^\alpha$ as representing four vector bosons,
i.e.\ massless `gravitons' with spin 1 and get a corresponding wave
equation $\quabla\, \theta_\mu^\alpha(x) = 0$ analogous to (\ref{KG}).
This could perhaps describe the gravitational field in non-standard form
(i.e.\ not as a spin-2 field) in the linearized limit. However, what we are
really looking for is a recipe to make this field dynamical and couple it
to matter. This was actually a main disadvantage of KL III.
The theory of the free Maxwell field followed from the fact that
$D(\frac{1}{2},0) \otimes D(\frac{1}{2},0) = D(\frac{1}{2},\frac{1}{2}) \oplus D(0,0)$
consists of a spin-1 triplett and a spin-0 singulett, where the former
can be written in terms of an anti-symmetric tensor with the algebraic
properties of Maxwell's field strength tensor. Even if we grant this
as a `derivation' of Maxwell's free theory, and even if we have also
obtained Dirac's equation as a free matter field theory, the
{\em dynamical coupling} of both fields is still missing.

Today, the usual recipe of coupling matter and interaction fields
is the gauge principle.
The idea is to start from a global symmetry of the free matter field theory
and then to postulate the corresponding local symmetry as well.
The geometrical picture is that of a fiber bundle over spacetime,
where the fibers are represented by the local symmetry group.
It thus turns out that the inhomogeneous term in the covariant derivative
(i.e.\ covariant under local gauge symmetry transformations)
has the interpretation of a bundle connection. A word of caution,
however: most textbooks present the connection coupling term
as a genuine way to introduce a real physical field.
But this is certainly an overestimation of a mere symmetry requirement!
The appearence of a connection only adjustes the postulate of local
gauge symmetry, the connection itself must still be considered flat,
i.e.\ with vanishing bundle curvature. A true physical interaction field
requires non-vanishing curvature and hence non-flat connections, but these
are clearly not \textit{enforced} by the logic of the gauge principle.
Local gauge transformations must be understood as a mere change
in the position representation of the wavefunction and are thus physically
vacuous. The situation is analogous to choosing curvilinear coordinates
in flat space (which leads to Christoffel symbols, i.e.\ connections,
but no gravitational field, i.e.\ non-vanishing curvature).
Therefore, one needs a true physical input to generally justify
the connection term as non-flat. In general relativity this input is
the equivalence principle. It has been argued that one possible way
to establish the connection term as a real coupling term could be
a \textit{generalization} of the equivalence principle
by introducing generalized `inertial' and `field charges'
(Lyre 2000). 
Whatever this physical input may be, in the following I will
understand the gauge principle already in such a stronger,
\textit{truly physical} sense.

Now, a gauge theory is characterized by a certain gauge group.
One nice starting point for an ur-theoretic gauge theory of gravity
could be the fact that the quantized ur-tetrad generates a group
which itself could be used as a proper gauge group. 
It can be shown that the Lie algebra of the ur-tetrad operators 
is 12-dimensional and that the corresponding Lie group is
isomorphic to $SL(2,\setC) \times SL(2,\setC)$.
Whether and how this group can be used for a proper gauge approach
is still an open question. One interesting point to be mentioned is
that the operators of this algebra ``live'' in flat Minkowski
spacetime (in ur-theory this is reflected by unimodular groups
rather than unitary ones).
As in many other gauge approaches of gravity we would therefore
describe the gravitational field in a flat space---and perhaps
in this sense it is indeed just a local field as other Yang-Mills fields.
But the point is that in ur-theory it will nevertheless
be possible to have a global spherical model of the cosmos,
since we got global curvature right from the beginning
by taking $\setS^3$ as the cosmic model.


\section{Tomorrow:
         Qubits, Holographic Principle and the Ontology of Information}
\label{tomorrow}

In this last section I shall come back to the more philosophical issues
of ur-theory. In Sect.\ \ref{yesterday} we already noticed that,
ontologically speaking, information in ur-theory seems to gain the status
of a modern notion of substance. Two questions are simply unavoidable then:
Does information exist without a carrier?
Does information exist without an observer or information-gathering system?

Again, the very idea of ur-theory is to characterize physical objects
entirely by the information which can be gained from them.
The further, novel feature is that even space or spacetime is
reconstructed here as a mere device to represent information.
By the time Weizs\"acker proposed it,
this was a revolutionary new perspective.
Nowadays, in modern quantum gravity, there is a strikingly similar
discussion about the deeper connections between space and information,
which has its roots in the considerations of black hole entropy.
In the early days, black holes were thought of as characterized
by three quantities only: mass, angular momentum and charge
(``no hair''-property). But this in turn means that we could
use them as `entropy graves' by simply feeding them with the
products of high entropy processes.
This caused Jakob Bekenstein (1973) to attribute
thermodynamical properties to black holes. 
In particluar, the surface of the event horizon $A$ in Planck units
turns out as a suitable measure of the entropy content
\be
\label{SBH1}
S = \frac{1}{4} A ,
\ee
thus leading to a generalized second law of thermodynamics.

Formula (\ref{SBH1}) is in many respects quite remarkable.
As Gerard 't Hooft (cf.\ his 2001) 
has pointed out, it does characterize physical objects
on the most fundamental level not by the three-dimensional volume
they occupy in space, but by a `projection' of their degrees
of freedom on a two-dimensional area much like a hologram.
He thus calls this idea the \textit{holographic principle}.
From an ur-theoretic point of view, it is rather the characterization
of physical objects in terms of pure information,
which becomes evident here. Indeed, if we calculate the
entropy and, hence, information content of the whole universe
in Planck lengths, we get $S_u \approx (10^{60})^2 = 10^{120}$ bits.
In the same way we get
$S_n \approx \left( \frac{\lambda}{l_o} \right)^2 = 10^{40}$
bits for a nucleon---hence, the ur-numbers Weizs\"acker already
found in the 60's! For an electron, however, (\ref{SBH1}) leads
to $S_e \approx \left( \frac{\lambda_e}{l_o} \right)^2 = 10^{46}$
in contrast to the $10^{37}$ urs stated in Sect.\ \ref{yesterday}.

Another calculation may suffice (cf.\ G\"ornitz 1988).
Suppose spherical symmetry $A=4 \pi R^2$ and the Schwarzschild
radius $R=2M$, then (\ref{SBH1}) transforms to
\be
\label{SBH2}
S = 4 \pi M^2 .
\ee
Again, for the whole universe $M_u = 10^{60} m_o$ we get
$S_u = 10^{120}$ bits as above. The information content of a particle
with mass $m$ then actually is the entropy difference of a universe
with or without such a particle, and hence
\be
\label{deltaS}
\Delta S = 4 \pi \Big( (M_u+m)^2 - M_u^2 \Big) \approx 8 \pi M_u m.
\ee
We now get $S_n \approx 10^{40}$ and $S_e \approx 10^{37}$
in accordance with the results from Sect.\ \ref{yesterday}.
It seems therefore that ur-theory does not directly support
the view of the holographic principle---and perhaps this
could be a helpful insight for other programs as well.

Indeed, the relevance of the Bekenstein entropy has been
acknowledged and partially explained in modern quantum gravity
programs---in string theory as well as in the quantum loop
approach (cf.\ Rovelli 1998b). 
Weizs\"acker's explanation---historically the first---is
a further alternative.
The Ashtekar-inspired, canonical quantum loop approach
has indeed some core similarities with ur-theory.
Like Penrose's twistor approach, quantum loop gravity
has its most suitable representation in terms of spin networks.
It is certainly the most powerful among the spinoristic programs
today, whose key feature is that they are background-free.
Of course, all of the programs mentioned are mathematically
by far more elaborated than ur-theory, which merely is an
outline or perhaps a raw framework of a spinoristic theory.
The true advantage of Weizs\"acker's approach is rather its
philosophical underpinning---and this may also serve other
programs.\footnote{For a recent dialogue between philosophy
of science and quantum gravity research see
Callender and Huggett (2001).}

To explore the implications of a true ontology of information
in physics, consider equation (\ref{SBH2}) again.
It indeed highlights the ur-theoretic view
that energy-matter \textit{is} information. In a future quantum
gravity, formula (\ref{SBH2}) might obtain the same status
as Einstein's $E=mc^2$, which indicates the ontological
equivalence of energy and matter. However, would this suffice
to consider information as a substance?
Surely, the age-old distinction between matter and form
lurks behind this question.
Aristotle considered a (physical) object a \textit{synholon}
that is a composition of form (\textit{eidos})
and matter (\textit{hyle}).
Form comes into matter, but none of these can exist independently,
as far as physical objects are concerned. For Aristotle
the essence of a thing is its form, the collection of all
the attributes characterizing the object completely.
In modern terms: the total information which can be gained from it.
Nevertheless Aristotle insisted in the necessity of matter.
He claimed the existence of a \textit{prote hyle}, a
``first matter'' as a kind of a neutral, universal substance.
The \textit{prote hyle} is nothing we can specify any further,
any form is stripped off from it.
It does therefore not belong to the realm of physics,
but can be considered the \textit{hypokeimenon} of
form---that on which form is based. Such a concept is
clearly a pure metaphysical one and whether we like to make
demands on it lies \textit{per definitionem} beyond physics.
What can empirically be known about an object,
is \textit{per se} reducible to information and in this sense
physics indeed reduces to information. Perhaps, ur-theory is
just the most rigorous anccount to take this insight seriously.

So far, information has been treated in an absolute sense.
However, from a conceptual point of view the notion of information
is inherently a context-related, relative concept presupposing
semantics (cf.\ Lyre 2002).    
As Weizs\"acker has put it: information does only exist in relation
to the difference of two semantical levels. E.g.\ the sign {\sf H}
on a sheet of paper has its `meaning' as the eighth letter
of the Latin capital alphabet or the seventh of the Greek;
but it may also mean just a collection of ink molecules.
To specify a particular amount of information, a certain semantics
must be presupposed, absolute numbers of information are `meaningless.'

From this point of view the above derived ur and, hence, bit numbers
appear as highly questionable. Under which ``context'' do they fall?
How could one specify the semantics of physical objects?
Actually, the idea of (context-) ``relatedness'' in physics is not
at all a new one, relativity theory heavily relies on it. In ur-theory,
the generators of the Poincar\'e group (cf.\ Castell 1975 and
Weizs\"acker 1985, p.~407) 
create and annihilate urs. Quite naturally, then,
the information content of a physical state becomes a frame-dependent
statement. This even applies to the information content of the universe.
We may think of the $10^{120}$ bits of the world or its age of
$10^{60}$ Planck times in relation to observers in the rest-frame
of the cosmos only. Proponents of certain quantum gravity approaches,
as for instance Carlo Rovelli (1997), 
advocate that only an extreme relationalist view of space and
time holds---as far as general covariance in general relativity
and its consequences in quantum gravity are concerned.
Ur-theory partially supports this view. On the other hand,
there is no other system left outside the universe.
Thus, the information content of the universe has an `absolute'
meaning in an operational sense. It is therefore not clear
to me whether it makes sense to tackle cosmology in a rigorous
generally covariant manner, where there is no distinguished
observer left. Operationally we \textit{are} distinguished as
inner observers of the universe with respect to its rest-frame.
In ur-theory the cosmic model is therefore not derived as a
solution of a certain set of basic equations, but as the epitome
of the existence of observers and physical objects in spacetime
itself: the representation of one ur.
Global cosmic spacetime should perhaps be treated significantly
different than local considerations.

Heisenberg's saying that there is no physics left, if we leave
aside the possibility of any observers at all, does apply here, too.
It was originally intended as a comment on the quantum measurement
problem, not on problems about global spacetime.
Ur-theory combines both meanings: urs are essentially quantum
wavefunctions of the cosmos---``cosmic qubits'' so to speak.
The relevance of the observer as a `device' whose factual
observation `creates' the observed thing is in ur-theory both
supported from a rigorous quantum as well as information theoretic
basis. As we already indicated in Sect.\ \ref{yesterday},
the ur as a qubit represents potential information.
Measurements must be understood as transitions from potential
to actual information. The upshot is that indeed the world only
comes into being for measuring observers---not for them as particular
individuals, but for the idea and possibility of such observers
in general.
Kant would have called this the \textit{transcendental subject}.
In my view it is the unavoidable epistemological consequence of
a fundamental physics of quantum information, and, eventually,
the philosophy of physics presumably has to deal with questions
of this sort.

Our speculations have come to a certain end now.
Perhaps I was able to support a bit the view that ur-theory is 
an account which allows for challenging considerations on
fundamental questions of physics and the philosophy of physics,
questions which are not even touched upon within the mainstream.
As a physical program, ur-theory surely suffers from mathematical
elaboration as compared to other programs and does therefore
not even earn the name ``theory'' in a strict sense.
But it is still highly impressing that Weizs\"acker had his main
starting ideas already at a time, when other programs were far
from being born. The main ingredients of ur-``theory''---information 
and spinorism---do exist as central motives in modern approaches 
as well (as pointed out in the above).
Perhaps the future development may lead to a mutual stimulation 
between other programs and the ur-hypothesis---at any rate this is 
what I wish Carl Friedrich von Weizs\"acker from the heart.


\vspace*{1cm}


\section*{References}

\begin{description}

\item[] 
Bekenstein, J.~D. (1973).
\newblock Black holes and entropy.
\newblock \textit{Physical Review D}, 7:2333--2346.

\item[] 
Callender, C. and Huggett, N., editors (2001).
\newblock \textit{Physics meets Philosophy at the Planck Scale}.
\newblock Cambridge University Press, Cambridge.

\item[] 
Castell, L. (1975).
\newblock Quantum Theory of Simple Alternatives.
\newblock In Castell, Drieschner, and Weizs{\"a}cker, editors, volume 1.

\item[] 
Castell, L., Drieschner, M. and Weizs{\"a}cker, C. F. v., editors (1975-1986).
\newblock \textit{Quantum {T}heory and the {S}tructures of {T}ime and {S}pace} (6 volumes).
\newblock Hanser, Munich.
  
\item[] 
Finkelstein, D. (1994).
\newblock Finite {P}hysics.
\newblock In Herken, R., editor,
\newblock \textit{The Universal Turing Machine - A Half-Century Survey}.
\newblock Springer, Vienna.

\item[] 
G{\"o}rnitz, T. (1988).
\newblock Abstract quantum theory and space-time-structure. {I}. {U}r
  theory and {B}ekenstein-{H}awking entropy.
\newblock \textit{International Journal of Theoretical Physics}, 27(5):527--542.

\item[] 
Lyre, H. (1998a).
\newblock \textit{Quantentheorie der Information}.
\newblock Springer, Vienna.

\item[] 
Lyre, H. (1998b).
\newblock Quantum space-time and tetrads.
\newblock \textit{International Journal of Theoretical Physics}, 37(1):393--400.
\newblock (E-print quant-ph/9703028).

\item[] 
Lyre, H. (2000).
\newblock A generalized equivalence principle.
\newblock \textit{International Journal of Modern Physics D}, 9(6):633--647.
\newblock (E-print gr-qc/0004054).

\item[] 
Lyre, H. (2002).
\newblock \textit{Informationstheorie. Eine philosophisch-naturwissenschaftliche
  Einf\"uhrung}.
\newblock Fink, Munich.
\newblock (UTB 2289).

\item[] 
Rovelli, C. (1997).
\newblock Halfway through the woods: Contemporary research on space and time.
\newblock In Earman, J. and Norton, J., editors,
\newblock {\em The Cosmos of Science}.
\newblock University of Pittsburgh Press/Universit{\"a}tsverlag Konstanz.

\item[] 
Rovelli, C. (1998).
\newblock Strings, loops and the others: A critical survey on the present
  approaches to quantum gravity.
\newblock In Dadhich, N. and Narlikar, J., editors, \textit{Gravitation and
  Relativity: At the Turn of the Millennium}. Poona University Press, Poona.
\newblock (E-print gr-qc/9803024).

\item[] 
't~Hooft, G. (2001).
\newblock Obstacles on the way towards the quantisation of space, time and
  matter---and possible resolutions.
\newblock \textit{Studies in History and Philosophy of Modern Physics},
  32:157--180.

\item[] 
Weizs{\"a}cker, C. F.~v. (1952).
\newblock Eine Frage \"uber die Rolle der quadratischen Metrik in der Physik.
\newblock \textit{Zeitschrift f{\"u}r Naturforschung}, 7 a:141.

\item[] 
Weizs{\"a}cker, C. F.~v. (1958).
\newblock Die {Q}uantentheorie der einfachen {A}lternative
  ({K}omplementarit{\"a}t und {L}ogik {I}{I}).
\newblock \textit{Zeitschrift f{\"u}r Naturforschung}, 13 a:245--253.

\item[] 
Weizs{\"a}cker, C. F.~v., Scheibe, E., and S{\"u}ssmann, G. (1958).
\newblock Komplementarit{\"a}t und {L}ogik, {I}{I}{I}. {M}ehrfache
  {Q}uantelung.
\newblock \textit{Zeitschrift f{\"u}r Naturforschung}, 13 a:705--721.

\item[] 
Weizs{\"a}cker, C. F.~v. (1985).
\newblock \textit{Aufbau der {P}hysik}.
\newblock Hanser, Munich.

\item[] 
Weizs{\"a}cker, C. F.~v. (1992).
\newblock \textit{Zeit und {W}issen}.
\newblock Hanser, Munich.

\end{description}


\end{document}